\documentclass[10pt,journal]{IEEEtran}
\usepackage[left=1.5cm, right=1.5cm, top=1.5cm, bottom = 1.5cm]{geometry}
\usepackage{amsmath}
\usepackage{amssymb}
\usepackage{amsthm}
\usepackage{graphicx}
\usepackage{mathtools}
\usepackage[ruled, vlined, linesnumbered]{algorithm2e}
\usepackage{color,soul}

\usepackage{cite}
\usepackage{subfigure}
\usepackage{stmaryrd}
\usepackage{breqn}
\usepackage{stackengine}
\usepackage{verbatim} 
\usepackage{glossaries}
\usepackage{color,soul}
\usepackage{xcolor}

\newcommand{\vectnorm}[1]{\left\lVert\begin{bmatrix}#1\end{bmatrix}\right\rVert_2}
\usepackage{xspace}

\newcommand{\nonl}{\renewcommand{\nl}{\let\nl\oldnl}}
\makeatletter
\newcommand\approxsim{\mathchoice
  {\@approxsim {\displaystyle}      {1ex} }
  {\@approxsim {\textstyle}         {1ex} }
  {\@approxsim {\scriptstyle}       {.7ex}}
  {\@approxsim {\scriptscriptstyle} {.5ex}}}
\newcommand\@approxsim[2]{%
  \mathrel{%
    \ooalign{%
      $\m@th#1\sim$\cr
      \hidewidth$\m@th#1.$\hidewidth\cr
      \hidewidth\raise #2 \hbox{$\m@th#1.$}\hidewidth\cr
    }%
  }%
}
\usepackage{etoolbox}

\let\mybibitem\bibitem
\renewcommand{\bibitem}[1]{%
  \ifstrequal{#1}{nature}
    {\color{black}\mybibitem{#1}}
    {\color{black}\mybibitem{#1}}%
}

\makeatother

\ifCLASSINFOpdf
 
\else
  
\fi

\begin{document}

\title{\LARGE On Optimizing the Power Allocation and the Decoding Order in Uplink Cooperative NOMA}
\author{Mohamed Elhattab,  Mohamed Amine Arfaoui, Chadi Assi, Ali Ghrayeb, and Marwa Qaraqe

\thanks{This paper was made possible by AICC03-0324-200005 from the Qatar National Research Fund (a member of Qatar Foundation). The findings herein reflect the work, and are solely the responsibility, of the authors.}}

\maketitle
\begin{abstract}
In this paper, we investigate for the first time the dynamic power allocation and decoding order at the base station (BS) of two-user uplink (UL) cooperative non-orthogonal multiple access (C-NOMA)-based cellular networks. In doing so, we formulate a joint optimization problem aiming at maximizing the minimum user achievable rate, which is a non-convex optimization problem and hard to be directly solved. To tackle this issue, an iterative algorithm based on successive convex approximation (SCA) is proposed. The numerical results reveal that the proposed scheme provides the superior performance in comparison with the traditional UL NOMA. In addition, we demonstrated that in UL C-NOMA, decoding the far NOMA user first at the BS provides the best performance.
\end{abstract}
\begin{IEEEkeywords}
C-NOMA, Fairness, Power allocation, Uplink.
\end{IEEEkeywords}
\vspace{-0.4cm}
\section{Introduction}
Non-orthogonal multiple access (NOMA) has been envisioned as a candidate multiple access technique for next-generation wireless networks \cite{Dai_A_2018}. The main idea of NOMA is to allow multiple user equipment (UEs) to access the same resource block (a time slot, a spreading code, or a frequency band), but with different power levels \cite{Dai_A_2018}. Specifically, NOMA utilizes superposition coding (SC) at the transmitter with proper power allocation as well as successive interference cancellation (SIC) at the receiver to cancel the inter-NOMA user interference \cite{Dai_A_2018}. It has been shown that NOMA outperforms the traditional orthogonal multiple access (OMA) techniques from the aspect of network connectivity and network spectral efficiency in both downlink (DL) and uplink (UL) transmissions \cite{Ali_Dynamic_2016}.  
\par In order to further improve the performance of NOMA cellular networks with low-cost infrastructure, the integration between NOMA and user-cooperative relaying referred to as cooperative NOMA (C-NOMA) has been developed \cite{CNOMA_User}. In C-NOMA, the near NOMA UEs, with typically good channel conditions, act as relays to assist the transmission of far NOMA UEs, who generally have bad channel conditions \cite{CNOMA_User}. As a result, by adding a cooperative link between a near NOMA UE and a far NOMA UE, a new degree of diversity alongside the base station (BS)-UEs link is introduced, thus allowing C-NOMA based wireless networks to achieve better fairness and higher spectral efficiency as compared to NOMA. \cite{CNOMA_User}.
\par Many work have investigated the performance of C-NOMA-based cellular networks and evaluated its performance in comparison with NOMA  in terms of minimum user achievable rate \cite{CNOMA_User}, network sum-rate \cite{Phuc_A_2020}, outage probability \cite{Elhattab_A_2020}, secrecy performance \cite{Chen_Physical_2018} and network power consumption \cite{Elhattab_Reconfigurable_2021}. Specifically, the authors in \cite{CNOMA_User} studied the power allocation to maximize the minimum UE achievable rate for a two-UE downlink C-NOMA network. The authors in \cite{Phuc_A_2020} discussed joint user pairing and power allocation to maximize the network sum-rate. The integration between coordinated multipoint transmission and C-NOMA was studied in \cite{Elhattab_A_2020} to improve the outage probability of the far NOMA UEs.  Finally, the integration between C-NOMA and reconfigurable intelligent surface was analyzed to minimize the network power consumption in \cite{Elhattab_Reconfigurable_2021}.  
\par It is worth mentioning that the aforementioned work and the citations therein solely investigate the potential gains of C-NOMA in the DL scenario, whereas research on its UL counterpart is still in its infancy stage. Recently, there has been a few work that focused on investigating the performance of UL C-NOMA networks \cite{Xie_2020_Ergodic, Liu_2019_Coordinated, Zhang_2018_Performance}. The authors in \cite{Xie_2020_Ergodic} and  \cite{Liu_2019_Coordinated} studied the performance analysis in terms of the ergodic capacity and the outage probability in a two-UE UL C-NOMA network, respectively, where \cite{Xie_2020_Ergodic} considered a dedicated full-duplex (FD) decode and forward (DF) relay; meanwhile \cite{Liu_2019_Coordinated} considered a dedicated half-duplex (HD) DF relay to assist the transmission of the two UL UEs. In addition, the authors in \cite{Xie_2020_Ergodic} and \cite{Liu_2019_Coordinated} assumed that there were no communication links between the BS and the UEs. In contrast, the authors in \cite{Zhang_2018_Performance} analyzed the performance of a two-UE UL C-NOMA in terms of the outage probability and the network sum-rate and considered a \textit{user cooperating relay} in which the near NOMA UE acted as a FD DF relay to assist the transmission from the far NOMA UE to the BS. However, all the aforementioned work studied only the performance analysis of an UL two-UE C-NOMA network while assuming a fixed non-optimal power control scheme. 
\par To the best of our knowledge, the dynamic power allocation in an UL two-UE C-NOMA network is yet to be investigated, thus motivating the study of this paper. Specifically, our main objective in this paper is to address the following questions.
\begin{itemize}
    \item \textbf{Q1:} What is the impact of employing a dynamic power allocation on the performance of UL C-NOMA over traditional UL NOMA?
    \item \textbf{Q2:} What is the appropriate decoding order in UL C-NOMA cellular networks?
    \item \textbf{Q3:} What are the main system parameters that emphasize the indispensability of invoking UL C-NOMA in cellular networks?
\end{itemize}

\indent In order to answer the above questions, we consider a two-UE UL C-NOMA system consisting of one BS, one near NOMA UE, denoted as UE$_{\rm{n}}$, and one far NOMA UE, denoted as UE$_{\rm{f}}$, in which UE$_{\rm{n}}$ acts a FD DF relay to assist the transmission from the UE$_{\rm{f}}$ to the BS. Specifically, we examine the impact of performing a dynamic power allocation and changing decoding order at the BS. In doing so, we formulate a power allocation problem to optimize the power allocation coefficients at UE$_{\rm{n}}$ and the transmit power at UE$_{\rm{f}}$ with the goal of maximizing the minimum UE's rate. Furthermore, we investigate two different SIC decoding orders at the BS, namely, far NOMA UE decoded first and near NOMA UE decodes first. The formulated optimization problem is a non-convex problem, which is difficult to be solved directly. To tackle this challenge, we propose a successive convex approximation (SCA)-based iterative algorithm to effectively solve the formulated optimization problem. We demonstrate the efficacy of the proposed algorithm by comparing its performance with the traditional UL NOMA. Our findings reveal that decoding the data of the far NOMA UE first always provides the best performance. Moreover, it is shown that the transmit power at far NOMA UE, the channel gain between the far and near NOMA UEs, and the self-interference channel have significant impact on the performance of UL C-NOMA. 
\begin{figure}
    \centering
    \includegraphics[width = 0.7 \columnwidth]{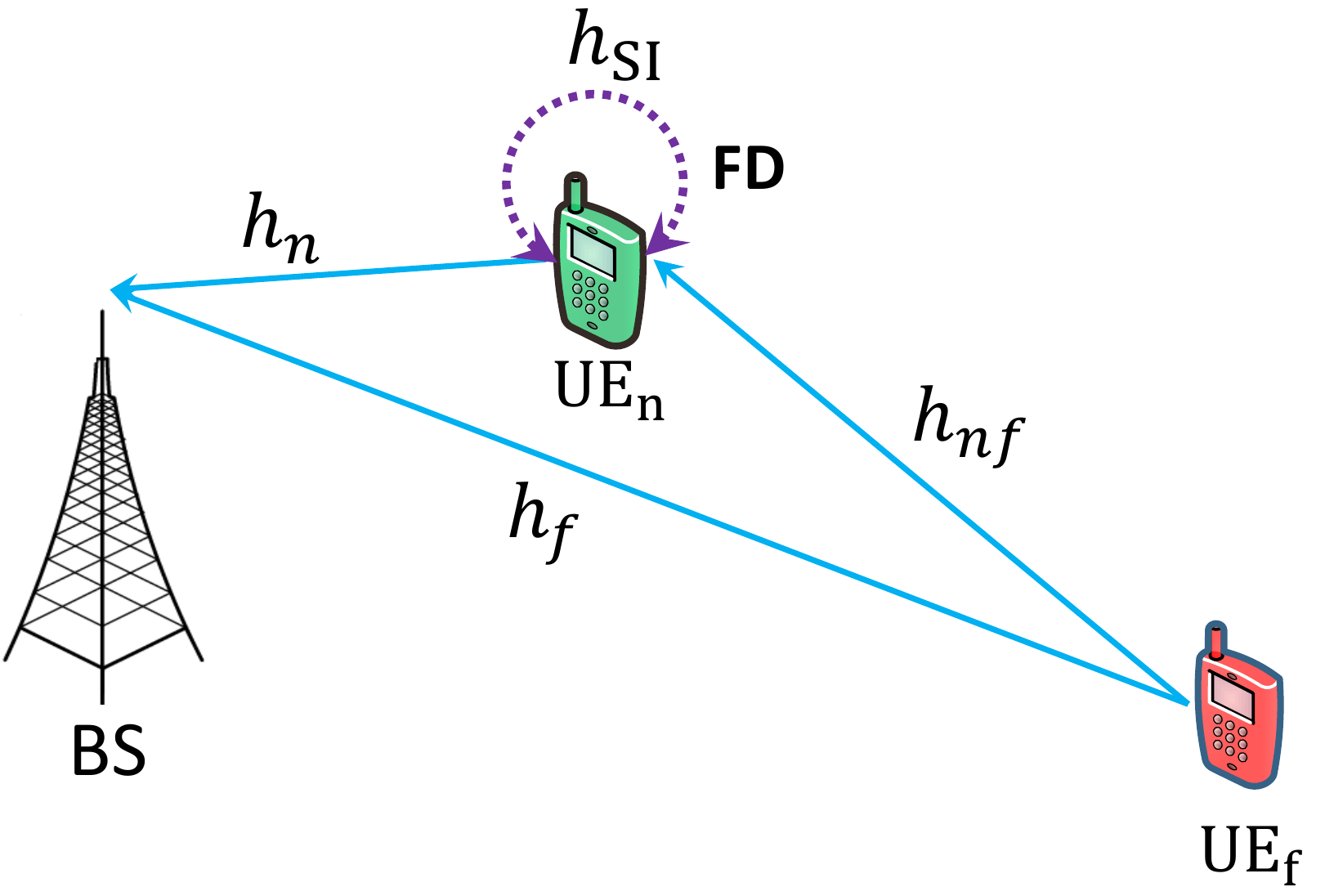}
    \caption{UL C-NOMA System Model}
    \label{fig: system}
\end{figure}
\vspace{-0.3cm}
\section{System, Transmission and Signal Models}
\subsection{Network Model}
\par We consider an UL transmission in a two-UE C-NOMA cellular network, which consists of one BS, one UE$_{\rm n}$ and one UE$_{\rm f}$ as shown in Fig. \ref{fig: system}. Similar to \cite{CNOMA_User, Simulation_setting}, we assume that the UEs and the BS each has one transmit and one receive antenna. Based on the NOMA principle, UE$_{\rm{n}}$ and UE$_{\rm{f}}$ can be served in the same resource (frequency/time) and form a NOMA pair. In order to enhance the performance of the traditional UL NOMA scheme, in this paper, we allow the UE$_{\rm{n}}$ to assist the transmission of UE$_{\rm{f}}$ by relaying the UE$_{\rm{f}}$'s message through a FD DF relaying mode. 
\par To this end, there are three wireless communication links, namely, UE$_{\rm{n}}$ $\rightarrow$ BS, UE$_{\rm{f}}$ $\rightarrow$ BS, and UE$_{\rm{f}}$ $\rightarrow$ UE$_{\rm{n}}$, whose channel coefficients are denoted, respectively, by $h_{n}, h_{f},$ and $h_{nf}$. In addition, there is a self-interference (SI) link arising from UE$_{\rm{n}}$ $\rightarrow$ UE$_{\rm{n}}$, whose channel coefficient is denoted by $h_{\rm{SI}}$. This SI is caused by the simultaneous transmission and reception at UE$_{\rm{n}}$. All wireless channels as well as the SI channel are independent and follow a Rayleigh distribution. Consequently, the channel gains for those wireless links follow Exponential distributions with parameters ${\lambda_{n}}$, ${\lambda_{f}}$, and ${\lambda_{nf}}$, respectively, and ${\lambda_{\rm SI}}$ for the SI link.  Finally, we denote the power budget of the UE$_{\rm{n}}$ and UE$_{\rm{f}}$ by $P_n^{\max}$ and $P_f^{\max}$, respectively. 
\vspace{-0.4cm}
\subsection{Transmission Model and Decoding Orders}
The transmission scheme in the considered model consists of two phases, described as follows.
\begin{itemize}
    \item \textit{First Phase:} UE$_{\rm{f}}$ broadcasts its own message, which is received by both the BS and the UE$_{\rm{n}}$. 
    \item \textit{Second Phase:} UE$_{\rm{n}}$ first decodes the message of UE$_{\rm{f}}$, then it superimposes its own message with the decoded message of UE$_{\rm{f}}$ using SC. Finally, UE$_{\rm{n}}$ transmits the superimposed signal to the BS.  
\end{itemize}

\indent Since UE$_{\rm{n}}$ adopts FD relaying mode, the two phases are executed within the same time-slot at the cost of the induced SI at the UE$_{\rm{n}}$  \cite{Phuc_A_2020}. Next, we discuss the possible decoding orders at the BS to detect the signals of UE$_{\rm{n}}$ and UE$_{\rm{f}}$. In this model, there are two possible decoding orders explained as follows.
\begin{enumerate}
     \item \textit{Far user decoded first (FUDF):} This decoding order is similar to the one applied in downlink NOMA/C-NOMA networks \cite{Ali_Dynamic_2016, Phuc_A_2020}. Specifically, the BS decodes the UE$_{\rm{f}}$'s message first considering the UE$_{\rm{n}}$'s signal as an interference signal. Then, the BS removes the signal of UE$_{\rm{f}}$ from its received signal to decode the message of UE$_{\rm{n}}$ without interference. 
    \item \textit{Near user decoded first (NUDF)}: This is the most widely used decoding order for UL NOMA transmission \cite{Ali_Dynamic_2016}. In this case, the BS first decodes the message of UE$_{\rm{n}}$ considering the UE$_{\rm{f}}$'s message as an interference signal. Subsequently, the BS decodes the message of UE$_{\rm{f}}$ free of interference. 
\end{enumerate}
\vspace{-0.4cm}
\subsection{Signal Model}
We start by describing the signal model at UE$_{\rm{n}}$ due to the transmission of UE$_{\rm{f}}$. Then, we discuss the signal model due to the transmission of both UE$_{\rm{n}}$ and UE$_{\rm{f}}$ at the BS. First, the received signal at UE$_{\rm{n}}$ in the $t$th time slot  due to the transmission of UE$_{\rm{f}}$ can be written as follows,
\begin{align}
    y_{n,1}  & = \left(\sqrt{\beta_f^{[m]} P_f^{\max}}h_{nf} x_f[t] \right. \notag \\ & \left. + \sqrt{(\alpha_n^{[m]} + \alpha_f^{[m]}) P_n^{\max}}h_{\rm{SI}} x_s[t-\tau] \right) + w_n[t],
\end{align}
where $x_s[t-\tau] = \alpha_n^{[m]} x_n + \alpha_f^{[m]} x_f$ is the superimposed signal at the UE$_{\rm{n}}$, $\alpha_n^{[m]}, \alpha_f^{[m]} \in [0, 1]$ are the power allocation coefficients for the $m$th decoding order that the UE$_{\rm{n}}$ assigns to transmit its own message and the message of UE$_{\rm{f}}$, respectively, and $\beta_f^{[m]} \in [0,1]$ is the fraction of power that UE$_{\rm{f}}$ utilizes from its power budget to transmit its own data. Moreover, $m \in \{1, 2\}$ represents the decoding order where $m=1$ indicates the decoding order is FUDF and $m=2$ indicates the decoding order is NUDF, $w_n$ is the additive white Gaussian noise with zero mean and variance of $\sigma^2$, and $\tau$ is the processing delay at UE$_{\rm{n}}$, which is assumed to be smaller than the time slot $t$ \cite{Phuc_A_2020, Zhang_2019_Resource}.  Based on this, the received signal-to-interference-plus-noise-ratio ($\tt{SINR}$) at UE$_{\rm{n}}$ to detect $x_f$ can be expressed as  
\begin{align}
    \delta_{n\longrightarrow f}^{[m]} = \frac{\beta_f^{[m]} P_f^{\max} \gamma_{nf}}{(\alpha_n^{[m]} + \alpha_f^{[m]}) P_n^{\max} \gamma_{\rm{SI}} + 1},
\end{align}
where $\gamma_{nf} = |h_{nf}|^2/\sigma^2$ and $\gamma_{\rm{SI}} = |h_{\rm{SI}}|^2/\sigma^2$. Consequently, the achievable rate at UE$_{\rm{n}}$ to decode the signal of UE$_{\rm{f}}$ can be expressed as $\mathcal{R}_{n \rightarrow f}^{[m]} = \log(1 + \delta_{n \rightarrow f}^{[m]}).$ 
On the other hand, in the second phase, UE$_{\rm{n}}$ superimposes the decoded signal of UE$_{\rm{f}}$, i.e. $x_f$, and its own message $x_n$, and then transmits this superimposed signal to the BS considering FD DF relaying mode. Consequently, at the end of the second phase, the received signal at the BS can be written as follows,
\begin{align}
    y_{b,2} &=  \underbrace{\sqrt{\alpha_n^{[m]} P_n^{\max}} x_n[t-\tau] + \sqrt{\alpha_f^{[m]} P_n^{\max}} x_f[t-\tau]}_{\text{Received signal from UE$_{\rm{n}}$ (second phase)}} \cr &  + \underbrace{\sqrt{\beta_f^{[m]} P_f^{\max}} h_{f} x_f[t]}_{\text{Received signal from UE$_{\rm{f}}$ (first phase)}} + w_b[t],
\end{align}
It is worth mentioning that the $\tt{SINR}$ and the achievable rate expressions for both UE$_{\rm{n}}$ and UE$_{\rm{f}}$ at the BS depend on the decoding order that the BS applies. Therefore, we discuss the $\tt{SINR}$ and rate expressions for both the FUDF and the NUDF schemes in the following section.
\vspace{-0.3cm}
\section{$\tt{SINR}$ and Rate Expressions}
\subsection{$\tt{SINR}$ and Rate Expressions for the FUDF Scheme}
In this scheme, the BS starts by decoding first the message of UE$_{\rm{f}}$ considering the transmission of UE$_{\rm{n}}$ to its own message as an interference signal. It is clear that the BS receives the message of UE$_{\rm{f}}$ from both UE$_{\rm{f}}$ and UE$_{\rm{n}}$.  As a result, at the end of the second phase, the BS employs the maximum ratio combining (MRC) technique to combine these receptions in order to decode the message of UE$_{\rm{f}}$ \cite{Phuc_A_2020}. Then, the interference signal caused by the signal of UE$_{\rm{f}}$ will be subtracted from the received signals using the SIC process to decode the message of UE$_{\rm{n}}$. Consequently, the $\tt{SINR}$ for decoding UE$_{\rm{f}}$'s message as well as the signal-to-noise-ratio ($\tt{SNR}$) for decoding the UE$_{\rm{n}}$'s message are given, respectively, by
\begin{align}
    \delta_{\rm{f}}^{[1]} &= \frac{\alpha_f^{[1]} P_n^{\max} \gamma_{n}}{\alpha_n^{[1]} P_n^{\max} \gamma_{n} + 1} + \beta_f^{[1]} P_f^{\max} \gamma_{f},\\
    \delta_{n}^{[1]}       &= \alpha_n^{[1]} P_n^{\max} \gamma_{n},
\end{align}
where  $\gamma_{f} = |h_{f}|^2/\sigma^2$ and $\gamma_{n} = |h_{n}|^2/\sigma^2$. Finally, the achievable data rate of UE$_{\rm{n}}$ as well as UE$_{\rm{f}}$ at the BS can be, respectively, expressed as 
\begin{align}
    \mathcal{R}_n^{[1]} = \log(1 + \delta_{{n}}^{[1]}),~\rm{and}~
    \mathcal{R}_f^{[1]} = \min(\mathcal{R}_{n \rightarrow f}^{[1]}, \mathcal{R}^{[1]}_{\rm{sum}}),
\end{align}
where $\mathcal{R}^{[1]}_{\mathrm{sum}} = \log_2(1 + \delta_f^{[1]})$. It is important to mention that since UE$_{\rm{n}}$ employs DF relaying, the achievable rate for UE$_{\rm{f}}$ is the minimum between $\mathcal{R}_{n \rightarrow f}^{[1]}$ and $\mathcal{R}^{[1]}_{\rm{sum}}$ \cite{Phuc_A_2020}.
\vspace{-0.3cm}
\subsection{$\tt{SINR}$ and Rate Expressions for the NUDF Scheme}
Different from the FUDF scheme, the UE$_{\rm{n}}$ is decoded first at the BS in the NUDF scheme, and hence, the received signal of the UE$_{\rm{f}}$ acts as an interference signal and its $\tt{SINR}$ at the BS to decode its own message can be expressed as
\begin{equation}
    \delta_{n}^{[2]} = \frac{\alpha_n^{[2]} P_n^{\max} \gamma_{n}}{\alpha_f^{[2]} P_n^{\max} \gamma_{n} + \beta_f^{[2]} P_f^{\max} \gamma_{f} + 1}.
\end{equation}
Based on this, the achievable data rate of UE$_{\rm{n}}$ at the BS can be expressed as $\mathcal{R}_n^{[2]} = \mathrm{log}(1 + \delta_n^{[2]})$. After the BS decodes the message of UE$_{\rm{n}}$, it removes it from the received signal to decode the message of UE$_{\rm{f}}$ without interference with the following $\tt{SNR}$
\begin{align}
    \delta_{f}^{[2]} &= \alpha_f^{[2]} P_n^{\max} \gamma_{n} + \beta_f^{[2]} P_f^{\max} \gamma_{f}.
\end{align}
Hence, the achievable rate of UE$_{\rm{f}}$ at the BS is expressed as 
\begin{equation}
    \mathcal{R}_f^{[2]} = \min (\mathcal{R}_{n \rightarrow f}^{[2]}, \mathcal{R}^{[2]}_{\mathrm{sum}}),
\end{equation}
where $\mathcal{R}^{[2]}_{\mathrm{sum}} = \log(1 + \delta_f^{[2]})$.
\vspace{-0.3cm}
\section{Proposed Power Control Scheme: Problem Formulation and Solution Approach}
\subsection{Problem Formulation}
In this paper, we emphasize the user fairness issue. Specifically, we investigate the joint optimization of the power allocation coefficients at UE$_{\rm{n}}$, i.e. $\alpha_n^{[m]}$ and $\alpha_f^{[m]}$, and the power allocation fraction at UE$_{\rm{f}}$, i.e., $\beta_f^{[m]}$, with the objective of maximizing the minimum user rate. The max-min rate optimization problem for a two-UE UL C-NOMA system is formulated as\footnote{In a multi-user scenario, an optimal UEs pairing policy should be first discussed to cluster one UE$_{\rm{n}}$ with one UE$_{\rm{f}}$. One can obtain the optimal pairing scheme using the Hungarian method \cite{Phuc_A_2020}. Then, the proposed power control scheme is applied for each NOMA pair to maximize the minimum user achievable rate in that pair. Note that, different NOMA pairs are served through orthogonal resources to avoid the inter-NOMA pair interference \cite{Phuc_A_2020}.} 
\allowdisplaybreaks
\begingroup
\begin{subequations}
\label{prob: Main}
\begin{align}
\mathrm{OPT}: &\quad \max_{{\alpha_n^{[m]}, \alpha_f^{[m]}, \beta_f^{[m]}}} \min(\mathcal{R}_f^{[m]}, \mathcal{R}_n^{[m]})  \\
\text{s.t.}\,\, & 0 \leq \alpha_n^{[m]} + \alpha_f^{[m]} \leq 1, \label{P1_C1}\\
&\quad \,\,\, 0 \leq \beta_f^{[m]} \leq 1, \label{P1_C2}
\end{align}
\end{subequations}
\endgroup
where constraints \eqref{P1_C1} and \eqref{P1_C2} ensure that the transmit powers at UE$_{\rm{n}}$ and at UE$_{\rm{f}}$ do not exceed their power budgets, i.e., $P_n^{\max}$ and $P_f^{\max}$, respectively. It can be seen that problem $\mathrm{OPT}$ is neither concave nor quasi-concave, which is difficult to be directly solved. In the next section, we present the solution approach when the adopted decoding order at the BS is FUDF, i.e. $m = 1$.\footnote{It is worth mentioning that the optimization problem $\mathrm{OPT}$ when $m = 2$ can be similarly solved by following the same steps in section IV-B.}
\vspace{-0.3cm}
\subsection{Solution Approach}
By introducing an auxiliary variable $\zeta$, the power control optimization problem can be rewritten as follows
\begingroup
\begin{subequations}
\label{prob: Main_1}
\begin{align}
\mathcal{P}: &\quad \max_{{\alpha_n^{[1]}, \alpha_f^{[1]}, \beta_f^{[1]}}} \zeta  \\
\text{s.t.}\,\, & \qquad  \eqref{P1_C1}, \eqref{P1_C2}, \\
\,\, & \log\left(1 + \alpha_n^{[1]} P_n^{\max} \gamma_{n}\right) \geq \zeta, \label{P2_C1} \\
\,\, & \log\left(1 + \frac{\alpha_f^{[1]} P_n^{\max} \gamma_{n}}{\alpha_n^{[1]} P_n^{\max} \gamma_{n} + 1} + \beta_f^{[1]} P_f^{\max} \gamma_{f}\right) \geq \zeta, \label{P2_C2}\\
\,\, & \log\left(1 + \frac{\beta_f^{[1]} P_f^{\max} \gamma_{nf}}{(\alpha_n^{[1]} + \alpha_f^{[1]}) P_n^{\max} \gamma_{\rm{SI}} + 1}\right) \geq \zeta. \label{P2_C3}
\end{align}
\end{subequations}
\endgroup
It can be observed that problem $\mathcal{P}$ is a non-convex optimization problem due to the non-convex constraints \eqref{P2_C2} and \eqref{P2_C3}. In order to tackle this challenge, we first reformulate constraints \eqref{P2_C2} and \eqref{P2_C3}, respectively, as follows.
\begin{align}
    \alpha_f^{[1]} P_n^{\max} \gamma_{n} + & \alpha_n^{[1]} \beta_f^{[1]} P_n^{\max}  P_f^{\max} \gamma_{n} \gamma_{f} + \beta_f^{[1]} P_f^{\max} \gamma_{f} \geq \notag \\ &\vartheta \alpha_n^{[1]} P_n^{\max} \gamma_{n} + \vartheta, \label{P2_C2_Mod}\\
    \beta_f^{[1]} P_f^{\max} \gamma_{nf} & \geq (\alpha_n^{[1]}\vartheta + \alpha_f^{[1]}\vartheta) P_n^{\max} \gamma_{\rm{SI}} + \vartheta, \label{P2_C3_Mod}
\end{align}
where $\vartheta$ is an auxiliary variable that should achieve the condition $\vartheta \geq \exp(\zeta) - 1$. Nevertheless, both \eqref{P2_C2_Mod} and \eqref{P2_C3_Mod} are intractable due to the multiplication of the two variables that exists in the left and right sides of those constraints. According to \cite{Xu_2017_Joint}, in order to tackle the multiplication of the variables in the right hand side of those constraints, and for any non-negative variables $x, y$, and $z$, the approximation of the following expression $xy \leq z$ can be formulated as $2xy \leq (ax)^2 + (y/a)^2 \leq 2z$, where the first inequality holds if and only if $a = \sqrt{y/x}$. Based on this, equations \eqref{P2_C2_Mod} and \eqref{P2_C3_Mod} can be rewritten as follows
\begin{align}
    \alpha_f^{[1]} P_n^{\max} \gamma_{n} + & \alpha_n^{[1]} \beta_f^{[1]} P_n^{\max}  P_f^{\max} \gamma_{n} \gamma_{f} + \beta_f^{[1]} P_f^{\max} \gamma_{f} \geq \notag \\ & v P_n^{\max} \gamma_{n} + \vartheta, \label{P2_C2_Mod_1}\\
    \beta_f^{[1]} P_f^{\max} \gamma_{nf} & \geq (v + u) P_n^{\max} \gamma_{\rm{SI}} + \vartheta, \label{P2_C3_Mod_1}
\end{align}
where the auxiliary variables $u$ and $v$ should satisfy the following constraints
\begin{align}
    \left(\frac{\alpha_n^{[1]}}{b^{[r]}}\right)^2 + \left(\vartheta b^{[r]}\right)^2 \leq 2 v,~\text{and}~\left(\frac{\alpha_f^{[1]}}{a^{[r]}}\right)^2 + \left(\vartheta a^{[r]}\right)^2 \leq 2 u,\label{eq: u}
\end{align}
such that $b^{[r]}$ and $a^{[r]}$ denote the values of $b$ and $a$ in the $r$th iteration, which can be updated by 
\begin{align}
    b^{[r]} = \sqrt{\frac{\alpha_n^{[1,r]}}{\vartheta^{[r]}}}~~\mathrm{and}~~ a^{[r]} = \sqrt{\frac{\alpha_f^{[1,r]}}{\vartheta^{[r]}}}, \label{Eq: Updat}
\end{align}
After the previous approximation, one can see that \eqref{P2_C3_Mod_1} is a convex constraint. However, \eqref{P2_C2_Mod_1} is still non-convex due to the variables multiplication in the left hand side. In order to handle this challenge, we introduce an auxiliary variable $\Lambda$ and equivalently define the following constraints
\begin{align}
    v P_n^{\max} \gamma_{n} + \vartheta &- \alpha_f^{[1]} P_n^{\max} \gamma_{n} -  \Lambda^2 P_n^{\max}  P_f^{\max} \gamma_{n} \gamma_{f} \notag \\ & - \beta_f^{[1]} P_f^{\max} \gamma_{f} \leq 0 \label{P2_C2_Mod_2} \\
    \alpha_n^{[1]} \beta_f^{[1]} & \geq \Lambda^2. \label{P2_C2_Mod_3}
\end{align}
Here, \eqref{P2_C2_Mod_3} is a quadratic conic convex constraint, meanwhile \eqref{P2_C2_Mod_2} is non-convex due to the concave function $f(\Lambda) = -\Lambda^2$ which renders the right side of \eqref{P2_C2_Mod_2} as a difference-of-convex (DC) form. Thus, with $\Lambda^{[r]}$ as the input point, we can apply the SCA technique to replace $f(\Lambda)$ by its first order Taylor approximate as:
\begin{align}
\tilde{f}(\Lambda;\Lambda^{[r])}) = - (\Lambda^{[r]})^2 - 2\Lambda^{[r]}(\Lambda-\Lambda^{[r]}). \label{approx:2}
\end{align}
At this point, problem $\mathcal{P}$ can then be replaced by:
\begingroup
\begin{subequations}
\label{prob: Main_2}
\begin{align}
\mathcal{P}_1: &\quad \max_{\substack{\alpha_n^{[1]}, \alpha_f^{[1]}, \beta_f^{[1]}, v, \\ \vartheta, \zeta, \Lambda, u}} \zeta  \\
\text{s.t.}\,\, & \qquad  \eqref{P1_C1}, \eqref{P1_C2} ,\eqref{P2_C3_Mod_1}, \eqref{eq: u}, \eqref{P2_C2_Mod_3},\\
\,\, & \qquad \vartheta \geq \exp(\zeta) - 1, \label{Eq: Cone}\\
\,\, & \qquad  \alpha_n^{[1]} P_n^{\max} \gamma_{n} \geq \vartheta, \label{Eq: P4_C2}\\ 
\,\, &  \qquad v P_n^{\max} \gamma_{n} + \vartheta + P_n^{\max}  P_f^{\max} \gamma_{n} \gamma_{f} \tilde{f}(\Lambda;\Lambda^{[m])}) \notag \\ &   \qquad - \alpha_f P_n^{\max} \gamma_{n} - \beta_f P_f^{\max} \gamma_{f} \leq 0. \label{Eq: P4_C3}
\end{align}
\end{subequations}
\endgroup
\begin{algorithm}[!t]
\label{Algorithm PS}
\DontPrintSemicolon
\small{
\caption{\small{Proposed Algorithm}}
\textbf{Input}{:~ channel gains $h_n, h_f, h_{n,f}, h_{\rm{SI}}$, $\sigma^2$, $P_f^{\max}, P_n^{\max}$, decoding order $m$, maximum number of iteration $N,$, and maximum tolerance $\epsilon = 10^{-4}$}; \;
\textbf{Initialize}{:~ Iteration index $r = 1$, $b^{[0]}, a^{[0]},$ and $\Lambda^{[0]}$};\;
\While {$r \leq N~\mathrm{and}~\zeta^{[r + 1]} - \zeta^{[r]} > \epsilon$}{ 
Increment $r := r + 1$;\;
Obtain the values of $\alpha_n^{[m,r]}, \alpha_f^{[m,r]}, \beta_f^{[m,r]}, \vartheta^{[r]}, \zeta^{[r]}, u^{[r]},$ and $v^{[r]}$ by solving problem $\mathcal{P}_2$;\;
Update $\Lambda^{[r]}$; \;
Update  $a^{[r]},$ and $b^{[r]}$ based on \eqref{Eq: Updat};}
\textbf{Output}:~ $\alpha_n^{[1]}, \alpha_f^{[1]}, \beta_f^{[1]},$ and $\zeta$;}
\end{algorithm}
One can remark that the obtained problem $\mathcal{P}_1$ is a generalized convex problem due to the existence of the generalized exponential cone constraints \eqref{Eq: Cone}. Although $\mathcal{P}_1$ can be efficiently solved using a convex solver \cite{Elhaber_2019_Joint}, it generally entails more computational time in comparison with other standard convex programs such as second order cone programming (SOCP) \cite{Elhaber_2019_Joint}. As a result, a conic programming solver may be used to provide a much more efficient practical implementation while achieving an accuracy of $99.99\%$ \cite{Elhaber_2019_Joint}. Consequently, this motivates us to invoke the conic approximation with controlled accuracy  in which constraint \eqref{Eq: Cone} can be rewritten by a set of second order cone inequalities as \cite{Elhaber_2019_Joint}:
\begin{align}
\label{prFc:main} \kappa_{q+4} & \leq 1 + \vartheta \\
1 + \kappa_1 & \geq \vectnorm{1 - \kappa_1 & 2 + \zeta/ 2^{q-1}} \notag\\
1 + \kappa_2 & \geq \vectnorm{1 - \kappa_2 & 5/3 + \zeta/ 2^q} \notag\\
\label{Eq: approx} 1 + \kappa_3 & \geq \vectnorm{1 - \kappa_3 & 2\kappa_1}\\
\kappa_4 & \geq \kappa_2 + \kappa_3/24 + 19/72 \notag\\
1 + \kappa_l & \geq \vectnorm{1 - \kappa_l & 2\kappa_{l-1}} \forall \ l \in \{5, ..., q+3\} \notag\\
1 + \kappa_{q+4} & \geq \vectnorm{1 - \kappa_{q+4} & 2\kappa_{q+3}} \notag 
\end{align}
\begin{figure*}[!t]
\centering
\hspace{-2cm}\subfigure[Transmit power at UE$_{\rm{n}}$] {\centering\includegraphics[width=0.28\textwidth]{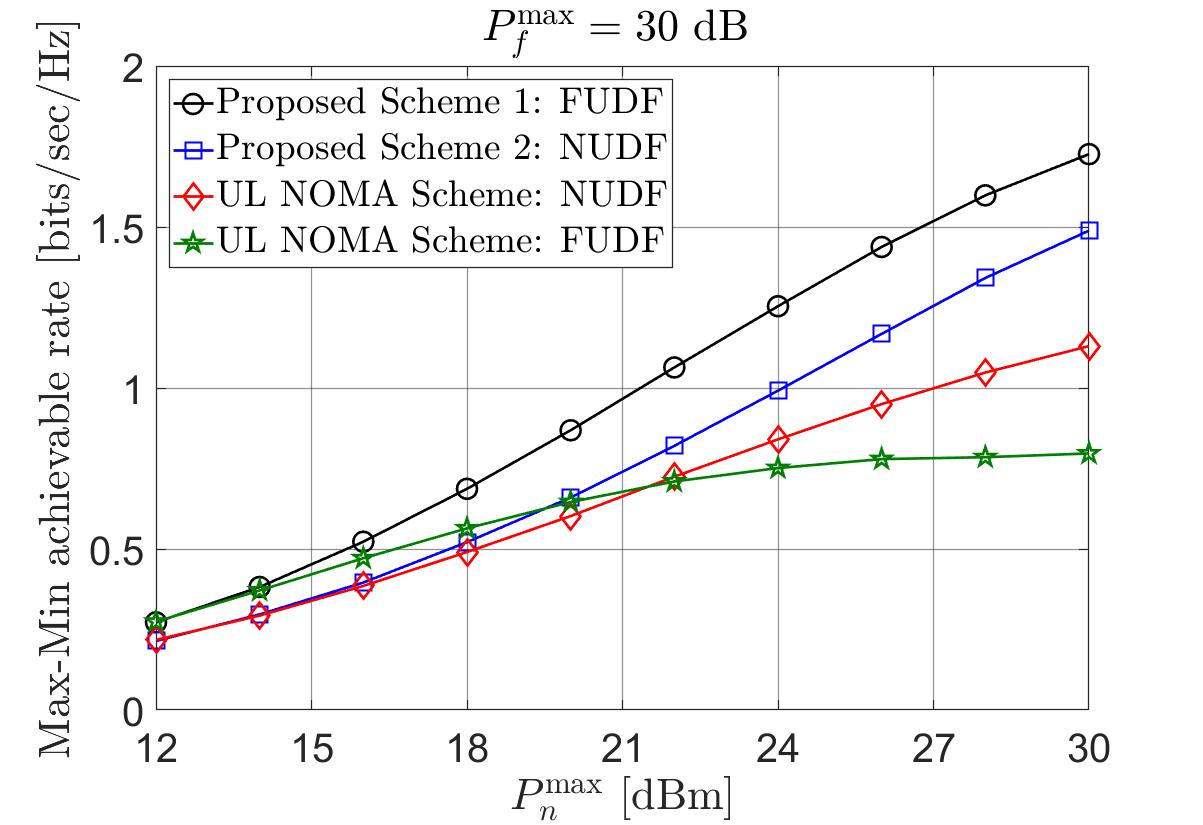}} 
\hspace{-0.5cm}\subfigure[Transmit power at UE$_{\rm{f}}$] {\centering\includegraphics[width=0.28\textwidth]{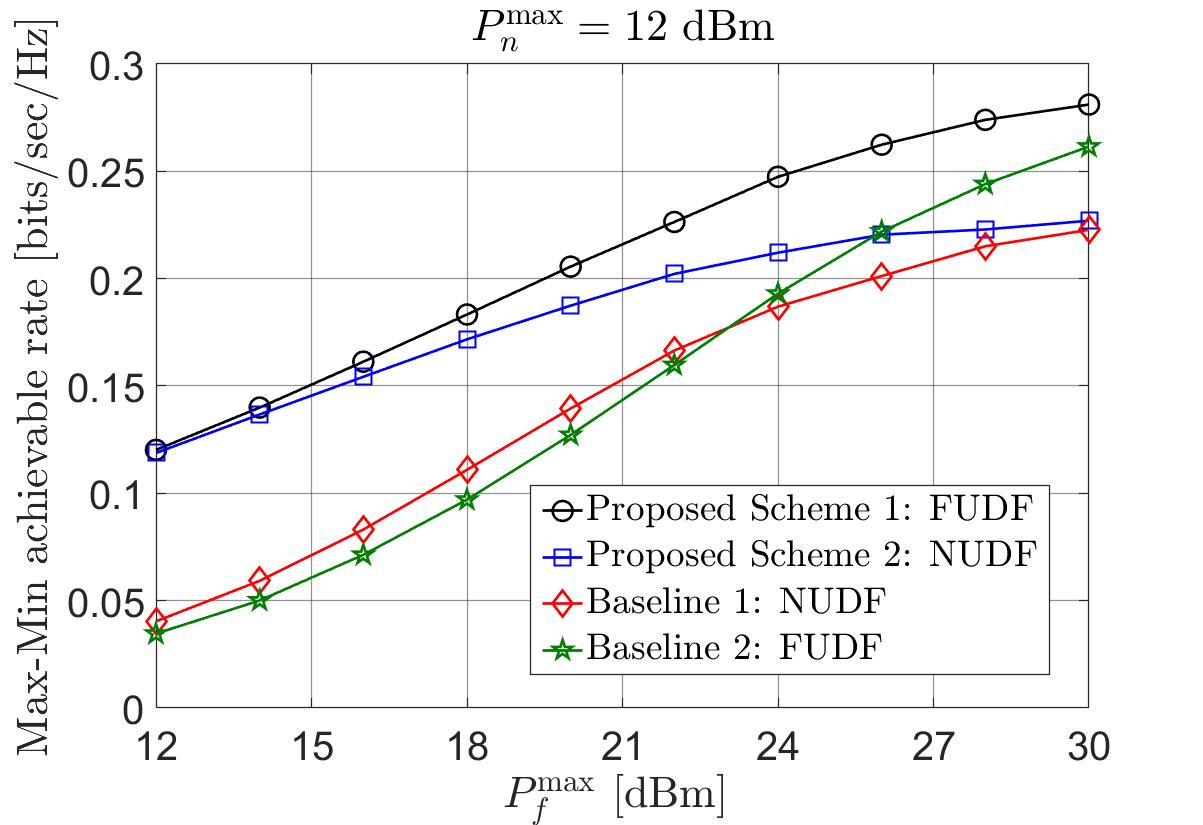}}  \hspace{-0.5cm}\subfigure[Self-Interference channel gain]
{\centering\includegraphics[width=0.28\textwidth]{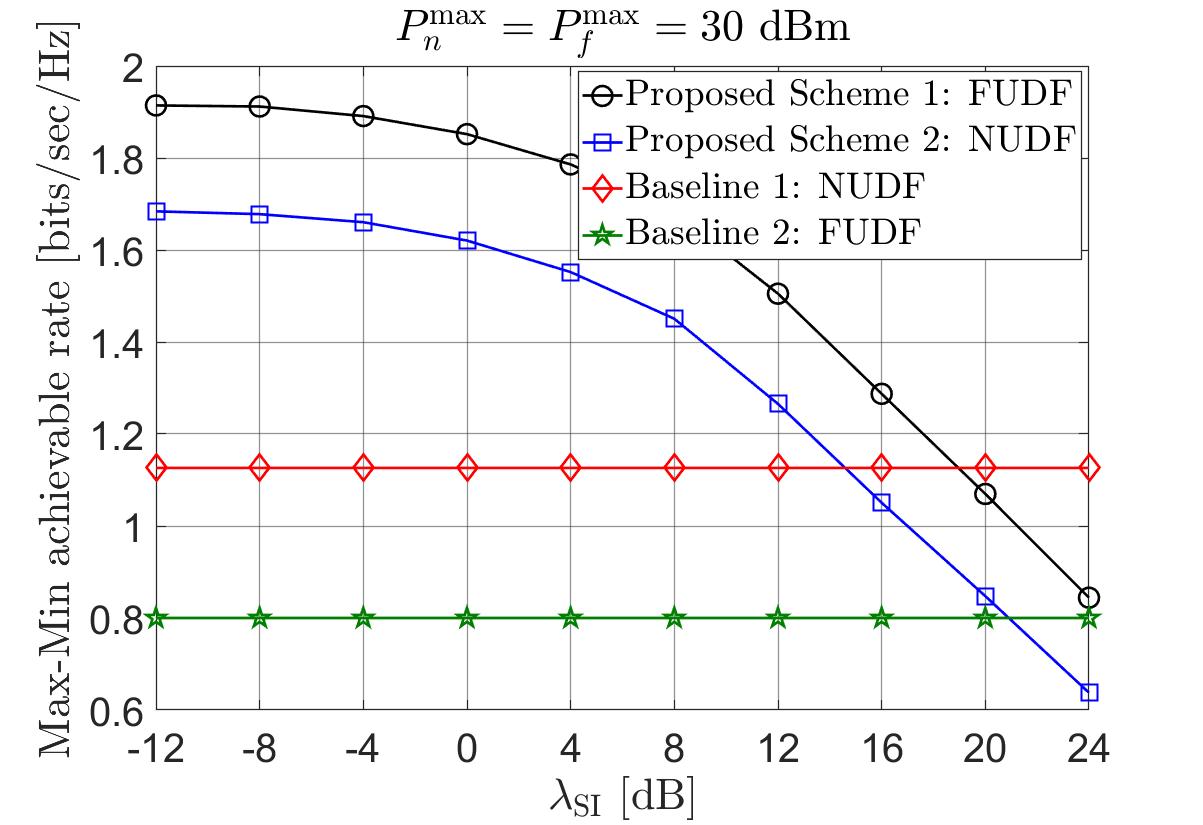}}
\hspace{-0.6cm} \subfigure[channel gain between UE$_{\rm{f}}$ and UE$_{\rm{n}}$]
{\centering\includegraphics[width=0.28\textwidth]{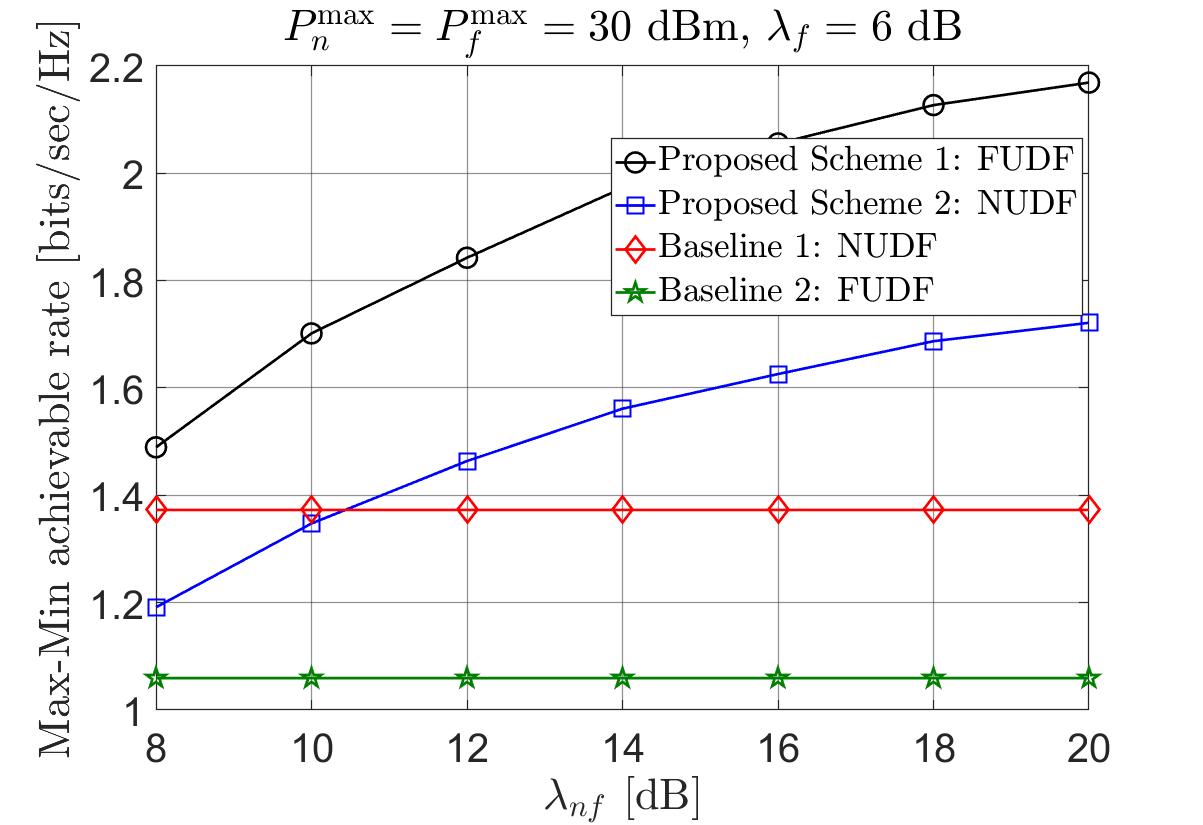}}\hspace{-2cm}
\caption{Performance evaluation of the proposed UL C-NOMA transmission.}
\label{fig: Simulation}
\vspace{-0.5cm}
\end{figure*}
where ${\kappa}_q, \forall q \in \{0,1, \dots, q+4\}$ is a new slack variable and $q$ is the parameter of the conic approximation technique to control the accuracy of the approximation, which can be chosen, according to \cite{Elhaber_2019_Joint}, as $q = 4$ to achieve around 99.99\% accuracy. Towards this end, problem $\mathcal{P}_1$ can be rewritten as 
\begingroup
\begin{subequations}
\label{prob: Main_3}
\begin{align}
&\hspace{-0.5cm}\mathcal{P}_2: \quad \max_{\substack{\alpha_n^{[1]}, \alpha_f^{[1]}, \beta_f^{[1]}, v, \\ \vartheta, \zeta, \Lambda, u, \kappa_q}} \zeta  \\
\text{s.t.}\,\, & \eqref{P1_C1}, \eqref{P1_C2} ,\eqref{P2_C3_Mod_1},\eqref{eq: u}, \eqref{P2_C2_Mod_3}, \eqref{Eq: P4_C2}, \eqref{Eq: P4_C3}, \eqref{prFc:main},~\text{and}~\eqref{Eq: approx}. \notag
\end{align}
\end{subequations}
\endgroup
Based on the above analysis, we can see that all the constraints in problem $\mathcal{P}_2$ are either linear or quadratic conic convex constraints. Hence, it can be efficiently solved using a standard optimization solver such as MOSEK \cite{Elhaber_2019_Joint}. Finally, the proposed iterative algorithm is presented in \textbf{Algorithm} 1, where  in each iteration, we solve $\mathcal{P}_2$ for given values of $\Lambda^{[r]}, a^{[r]},$ and $b^{[r]}$ to obtain the optimal values of  $\alpha_n^{[m]}, \alpha_f^{[m]}, \beta_f^{[m]}, \vartheta, \zeta, u,$ and $v$. Then, we update the iteration index $r$ and the parameters $\Lambda^{[r]}, a^{[r]},$ and $b^{[r]}$ to solve $\mathcal{P}_2$ in the next iteration until the condition either $r > N$ or $\left(\zeta^{[r + 1]} - \zeta^{[r]}\right) \leq \epsilon$ is satisfied.
\vspace{-0.3cm}
\section{Simulation Results and Discussion}
In this section, our objective is to analyze the performance
of the two proposed decoding schemes, FUDF and NUDF, under various
system parameters by varying the power budget at UE$_{\rm{n}}$, the power budget at UE$_{\rm{f}}$, the SI channel gain, and the channel gain between UE$_{\rm{n}}$ and UE$_{\rm{f}}$. The simulation results are obtained through generating $10^3$ independent Monte-Carlo trials. Unless otherwise mentioned, we adopt the same simulation setting in \cite{CNOMA_User}, in which $\lambda_n = \lambda_{nf} = 12$ dB, $\lambda_f = 3$ dB, and $\lambda_{\rm{SI}} = 5$ dB. In order to highlight the effectiveness of the proposed schemes, we compare them with the two following baselines that are based on the traditional UL NOMA scheme \cite{Wei_2017_Fairness, Ali_Dynamic_2016}.
\begin{enumerate}
    \item \textit{UL NOMA considering NUDF approach}: This scheme is denoted as baseline 1 in which UE$_{\rm{n}}$ is decoded first at the BS.
    \item \textit{UL NOMA considering FUDF approach}: This scheme is denoted as baseline 2 in which UE$_{\rm{f}}$ is decoded first at the BS.
\end{enumerate}
First, one can see from Fig. \ref{fig: Simulation} that the proposed FUDF scheme achieves the best performance compared to the other proposed NUDF scheme as well as the two considered baselines schemes. This observation is different from the traditional UL NOMA scheme \cite{Ali_Dynamic_2016}, where, in the majority of the cases, decoding the near user first (baseline 1) is better than decoding the far user first (baseline 2). This advises that the BS should always decode the far user first in UL C-NOMA networks. 
\par Fig. \ref{fig: Simulation}(a) depicts the effect of increasing the power budget at UE$_{\rm{n}}$ on the performance of the four schemes. It can be seen that at a low power budget at UE$_{\rm{n}}$ and a high power budget at UE$_{\rm{f}}$, the FUDF scheme almost achieves the same performance as baseline 2. This is because most of the power of the UE$_{\rm{n}}$ is assigned to its own message rather than UE$_{\rm{f}}$'s message in order to improve its performance, since the high power budget at UE$_{\rm{f}}$ substitutes its weak channel with the BS. However, when $P_n^{\max}$ increases, the proposed FUDF achieves a higher performance gain compared to the two baselines schemes. This is because increasing $P_n^{\max}$ motivates UE$_{\rm{n}}$ to assist UE$_{\rm{f}}$ and, hence, improves the performance of both UE$_{\rm{n}}$ and UE$_{\rm{f}}$.
\par Fig. \ref{fig: Simulation}(b) presents the effect of $P_f^{\max}$ on the system performance. Note that, when $P_f^{\max}$ is low, the cooperation between UE$_{\rm{n}}$ and UE$_{\rm{f}}$ is indispensable. The main reason behind this is that the received signal at the BS due to the transmission of UE$_{\rm{f}}$ is weak and UE$_{\rm{n}}$ should assist the transmission to improve UE$_{\rm{f}}$'s achievable rate. On the other hand, when $P_f^{\max}$ increases, and since $P_n^{\max}$ is low, UE$_{\rm{n}}$ starts to allocate most of its power to its own message and, hence, the performance gains due to the cooperation diminishes. Fig. \ref{fig: Simulation}(c) shows the effect of SI channel gain on the C-NOMA schemes. Increasing the SI channel gain forces UE$_{\rm{n}}$ to reduce its transmit power so that it can avoid harming itself. As a result, when the SI values is relatively large, UE$_{\rm{n}}$ transmits with very low power and the performance of the UL C-NOMA becomes worse than that of UL NOMA. Finally, Fig. \ref{fig: Simulation}(d) presents the effect of $\lambda_{nf}$ on the system performance. It can be seen that the performance gain between C-NOMA and NOMA when the cooperative link has a bad channel condition is low. This is reasonable because C-NOMA mainly depends on the cooperative link. Meanwhile, when $\lambda_{nf}$ is relatively good, the performance of the DF relaying enhances and, hence, the minimum user rate improves.  
\vspace{-0.3cm}
\section{Conclusion}
In this paper, we investigated the performance of the UL C-NOMA and compared its performance with the traditional UL NOMA. In contrast to UL NOMA technique, we have shown that the BS should always decode first the far user's message. In addition, we have also observed that UL C-NOMA achieves a superior performance compared to UL NOMA, especially with low power budget at far NOMA user, moderate values for the SI channel, and with good channel conditions between the far NOMA user and the near NOMA user.   
\vspace{-0.3cm}
\bibliographystyle{IEEEtran}
\bibliography{IEEEabrv,reference}

\end{document}